\documentclass[twocolumn,letter]{jpsj3}
\usepackage{graphicx}
\usepackage{bm}

\newcommand{\lihoyf}{LiHo$_{x}$Y$_{1-x}$F$_{4}$}
\newcommand{\thcrsi}{ThCr$_{2}$Si$_{2}$}
\newcommand{\dyrusi}{DyRu$_{2}$Si$_{2}$}
\newcommand{\yrusi}{YRu$_{2}$Si$_{2}$}

\newcommand{\dyyrusi}{Dy$_{x}$Y$_{1-x}$Ru$_{2}$Si$_{2}$}
\newcommand{\gdyrusi}{Gd$_{x}$Y$_{1-x}$Ru$_{2}$Si$_{2}$}
\newcommand{\femntio}{Fe$_{x}$Mn$_{1-x}$TiO$_{3}$}

\newcommand{\Tg}{T_{\mbox{\scriptsize g}}}

\newcommand{\chieq}{\chi_{\mbox{\scriptsize eq}}}
\newcommand{\chinl}{\chi_{\mbox{\scriptsize nl}}}
\newcommand{\Meq}{M_{\mbox{\scriptsize eq}}}
\newcommand{\Cmag}{C_{\mbox{\scriptsize mag}}}

\newcommand{\zMF}{z_{\mbox{\scriptsize MF}}}
\newcommand{\nuMF}{\nu_{\mbox{\scriptsize MF}}}
\newcommand{\znuMF}{(z\nu)_{\mbox{\scriptsize MF}}}

\title{Existence of a Phase Transition under Finite Magnetic Field \\ in the Long-Range RKKY Ising Spin Glass \dyyrusi }

\author{Yoshikazu Tabata\thanks{E-mail address: y.tabata@ht4.ecs.kyoto-u.ac.jp},  
Kousuke Matsuda, Satoshi Kanada, Teruo Yamazaki, Takeshi Waki, Hiroyuki Nakamura, 
$^{1}$Keisuke Sato and $^{1}$Koichi Kindo 
}
\inst{ Department of Materials Science and Engineering, Kyoto University, Kyoto 606-8501, Japan \\
$^{1}$The Institute for Solid State Physics, University of Tokyo, Kashiwa 277-8581, Japan
}
\abst{
 A phase transition of a model compound of the long-range Ising spin glass (SG) \dyyrusi , where spins interact via the Ruderman-Kittel-Kasuya-Yoshida (RKKY) interaction, has been investigated. The static and dynamic scaling analyses  reveal that the SG phase transition in the model magnet belongs to the mean-field universality class. Moreover, the characteristic relaxation time in finite magnetic fields, as well as in zero field, exhibits a critical divergent behavior, indicating the stability of the SG phase in finite fields. The presence of the SG phase transition in the field in this model magnet strongly suggests that the replica symmetry is broken in the long-range Ising SG.
 }

\kword{spin glass, critical phenomena, replica symmetry breaking, \dyyrusi }

\begin{document}
\maketitle

It has been a great challenge in the statistical physics to clarify the nature of glassy states. 
In this context, much attention has been paid to the spin glass (SG) \cite{SGreview} as the simplest example of glassy systems. According to the theoretical studies on the mean-field model of the SG, the Sherrington and Kirkpatrick (SK) model, the SG state is a thermodynamic equilibrium phase with the replica symmetry breaking (RSB) \cite{MFSG}.  The RSB SG state is essentially different from uniform states such as paramagnetic and (anti)ferromagnetic states where the replica symmetry (RS) is preserved. The RSB SG state has a complicated multi-valley structure of the free energy \cite{MezardRSB}, whereas the RS state has a simple one- or two-valley structure. 
The SK model can explain the glassy nature of the SG qualitatively, and the RSB is a vital concept for understanding the glassy state.

However, an application of the SK model to real SG systems is not valid as it is, because the interaction range is assumed to be infinite in the SK model, whereas it is finite in real systems. 
An alternative theory of the SG, the so-called droplet theory based on the short-range interaction model \cite{FisherHuseDroplet}, predicts no RSB and that only two thermodynamic states related to each other by a global spin flip exist in the SG state, as in uniform states. 
The nature of the SG state in real systems, the RSB state or the RS state, is still a controversial issue even at present.

The most striking difference between the RSB and RS SG states is the stability of the SG state under magnetic field. In the mean-field picture, a phase transition with the RSB occurs on the de Almeida-Thouless (AT) line under magnetic field \cite{AT}. On the other hand, there is no SG phase transition at finite temperature under magnetic field in the droplet picture \cite{FisherHuseDroplet}. 
Many numerical and experimental studies have been performed to examine the stability of the SG state under magnetic field in real materials. Detailed studies of an Ising SG with short-range interaction \femntio\ gave evidence against the SG transition in finite magnetic field \cite{NordbladFeMnTiO,JonssonFeMnTiO}. Recent numerical studies \cite{ShortIsingField} support these experimental results and claimed that the short-range Ising SG is the RS state described by the droplet picture \cite{LeuzziISGunderHAPPENDIX}. 

However, the nature of the SG state is still an open question in other classes, for instance, the long-range interaction system. The spacial interaction range, as well as the spacial dimension, is relevant to the phase transition, and hence, the long-range Ising SG could have a different nature from the short-range one. The aim of this study is to verify whether the thermodynamic SG state under magnetic field is preserved in long-range interaction systems and the possibility of the RSB state in real materials. Here, we report on recent experimental studies of a long-range RKKY Ising SG  \dyyrusi . The RKKY interaction via conduction electrons is a representative long-range interaction in real materials. Our results clearly indicate the presence of the SG phase in magnetic field and strongly suggest the emergence of the RSB in the long-range Ising SG, in contrast to the short-range Ising SG.

\begin{figure}[ht]
 \begin{center}
 \includegraphics[clip,width=8cm]{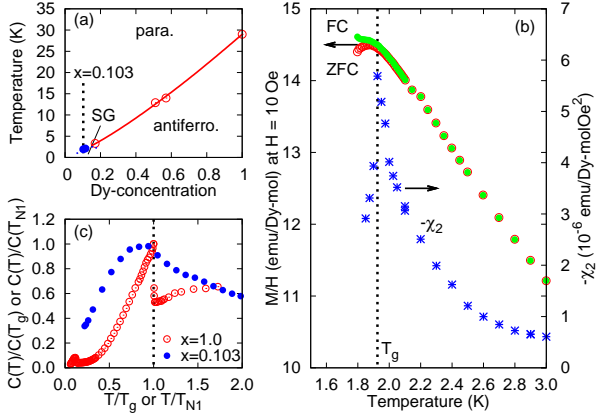}
 \end{center}
 \caption{ \label{fig1}(Color online) (a) Phase diagram of \dyyrusi . (b) $T$ dependences of $M/H$ with the ZFC and FC conditions and $-\chi_{2}$.  (c) $T$ dependences of $\Cmag$ of the $x$ $=$ 1.0 and 0.103 compounds. The horizontal and vertical axes are the scaled temperature and the scaled $\Cmag$, respectively. 
}
\end{figure}

\dyrusi\ is a rare-earth intermetallic compound with the tetragonal \thcrsi -type crystal structure and exhibits two successive antiferromagnetic transitions with $T_{\mbox{\scriptsize N}1}$ $=$ 29 K and $T_{\mbox{\scriptsize N}2}$ $=$ 3.5 K \cite{IwataDRS}. The compound has a strong uniaxial magnetic anisotropy and is recognized as an Ising antiferromagnet, where Dy$^{3+}$ spins interact via the long-range RKKY interaction. In the diluted compound \dyyrusi , a SG transition occurs with $x$ $=$ 0.103, as shown in Fig. \ref{fig1}(a). Consequently, \dyyrusi\ is a good model compound for investigating the nature of the long-range Ising SG. 

The single crystalline samples of \dyyrusi\  were grown by the Czhochralski method with a tetra-arc furnace. The concentration of Dy was determined by comparing the saturated magnetization of the diluted compound along the magnetic easy c-axis with that of the pure compound \dyrusi . Ac and dc magnetizations were measured  with the SQUID magnetometer MPMS (Quantum Design) equipped in the Research Center for Low Temperature and Material Sciences in Kyoto University. 
The dc magnetization was measured 10 minutes after stabilization at a certain temperature and a certain field because the samples showed a very slow dynamics, as described later. 
The ac measurements were performed with an ac field of 3 Oe and a frequency of 0.01 Hz $\leq$ $\omega$ $\leq$ 1000 Hz. 
We use a thin plate-shaped sample with a weight of 7.0 mg and a size of $5 \, \times 1 \, \times 0.2$ mm$^{3}$ for the dc and ac magnetization measurements to inhibit the effects of the diamagnetic field and the eddy current. 
Specific heat was measured with PPMS (Quantum Design). The lattice contribution of the specific heat of \dyyrusi\ was estimated from the specific heat of a non-magnetic reference compound \yrusi .

Figure \ref{fig1}(b) shows the temperature ($T$) dependence of the dc magnetization divided by the magnetic field $M/H$ at $H$ $=$ 10 Oe. The dc magnetization was measured under the zero-field-cooled (ZFC) and field-cooled (FC) conditions. A distinct separation of the ZFC and FC magnetizations is found around 1.9 K. In Fig. \ref{fig1}(b), the $T$ dependence of the second nonlinear susceptibility $\chi _{2}$ is also shown. $\chi_{2}$, which is the order parameter susceptibility of the SG, was obtained by fitting the $M(H,T)/H$ at each temperature as a function of $H^{2}$ in the form of $M(H,T)/H$ $=$ $\chi_{0}(T) + \chi_{2}(T) H^{2}  + \dotsb$. The fitting range of $H$ should be sufficiently low to extract $\chi_{2}$, and thus, $H$ $\leq$ 50 Oe was chosen to be around 2 K.  As shown in Fig. \ref{fig1}(b), $\chi_{2}$ exhibits a negative divergent behavior toward $\sim$ 1.9 K.  The separation of the ZFC and FC magnetizations and the negative divergent behavior of $\chi_{2}$ are characteristics of the SG phase transition in the zero field limit. Figure \ref{fig1}(c) shows the $T$ dependence of the magnetic specific heat ($\Cmag$) of the $x$ $=$ 0.103 compound with the data of the pure \dyrusi . In the plot, the horizontal axis is the scaled temperatures $T/\Tg$ and $T/T_{\mbox{\scriptsize N}1}$ for each compound. The SG transition temperature $\Tg$ of the $x$ $=$ 0.103 compound is obtained by the static scaling analysis described later. The anomaly of $\Cmag$ of the $x$ $=$ 0.103 compound at $\Tg$ is a cusp-like one, being much weaker than that of the pure compound at $T_{\mbox{\scriptsize N}1}$. The cusp-like anomaly of $\Cmag$ at $\Tg$ is similar to that expected by using the SK model; this is in contrast to general SG materials, which show no anomaly of $\Cmag$ at $\Tg$ \cite{SGreview}. 

\begin{figure}[ht]
 \begin{center}
 \includegraphics[clip,width=8cm]{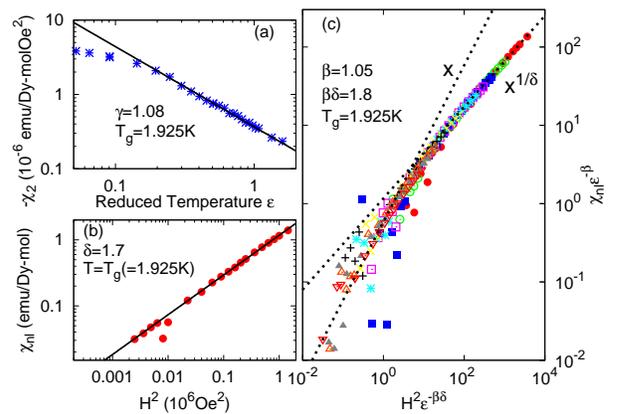}
 \end{center}
 \caption{ \label{fig2} (Color online) (a) Log-log plot of $-\chi _{2}$ vs $\varepsilon$. (b) Log-log plot of $\chinl$ vs $H^{2}$ at $T$ $=$ $\Tg$. (c) Full scaling plot  of $\chinl \varepsilon^{-\beta}$ vs $H^{2}\varepsilon^{-\beta\delta}$.  Dotted lines represent the asymptotic behaviors of the scaling function $F(x)$ as $x$ for $x$ $\rightarrow$ 0 and $x^{1/\delta}$ for $x$ $\rightarrow$ $\infty$.
}
\end{figure}

To examine the static critical behaviors of the SG phase transition in \dyyrusi , we analyzed the nonlinear susceptibility of the $x$ $=$ 0.103 compound in more detail. First, a log-log plot of $-\chi_{2}$ against the reduced temperature $\varepsilon$ $\equiv$ $(T/\Tg -1)$ is shown in Fig. \ref{fig2}(a). The best plot showing a linear relation between $\log (-\chi_{2})$ and $\log \varepsilon$ is obtained with $\Tg$ $=$ 1.925 K and the critical exponent $\gamma$ $=$ 1.08. As approaching $\Tg$ closer than $\varepsilon$ $<$ $0.1$, the log-log dependence of $\chi_{2}$ fails. This might be the effect of contamination of higher terms. Second, the $H^{2}$ dependence of the field-dependent nonlinear susceptibility $\chinl (H,T)$ $=$ $\chi_{0}(T) - M(H,T)/H$ at $\Tg$ $=$ 1.925 K is shown in Fig. \ref{fig2}(b) in the double logarithmic scale. $\chinl (H,T)$ corresponds to the SG order parameter and the experimental data exhibits a liner relation in Fig. \ref{fig2}(b) up to $H$ $=$ 1000 Oe with the critical exponent $\delta$ $=$ 1.7. Third, we attempted the scaling plot of $\chinl (H,T)$ in the form of $\chinl \varepsilon^{-\beta}$ vs $H^{2}\varepsilon ^{-\beta\delta}$. As shown in Fig. \ref{fig2}(c), the experimental data in the ranges of $\varepsilon$ $\leq$ 1.5 and $H^{2}$ $\leq$ $10^{6}$ Oe$^{2}$ collapse on a single curve with $\beta$ $=$ $1.05 \pm 0.15$ and $\beta\delta$ $=$ $1.8 \pm 0.1$, indicating that the nonlinear susceptibility exhibits the scaling form of 
\begin{equation}
 \chinl (T,H) = \varepsilon^{\beta} F\left(H^{2}\varepsilon^{-\beta\delta}\right) .
\end{equation}
The obtained set of critical exponents is roughly satisfied with the scaling relation of $\gamma+\beta$ $=$ $\beta\delta$ and is quite different from that in the short-range Ising SG \femntio\ \cite{GunnarssonFeMnTiO}. Instead, it is very close to that in the SK model \cite{SGreview}.

\begin{figure}[ht]
 \begin{center}
 \includegraphics[clip,width=8cm]{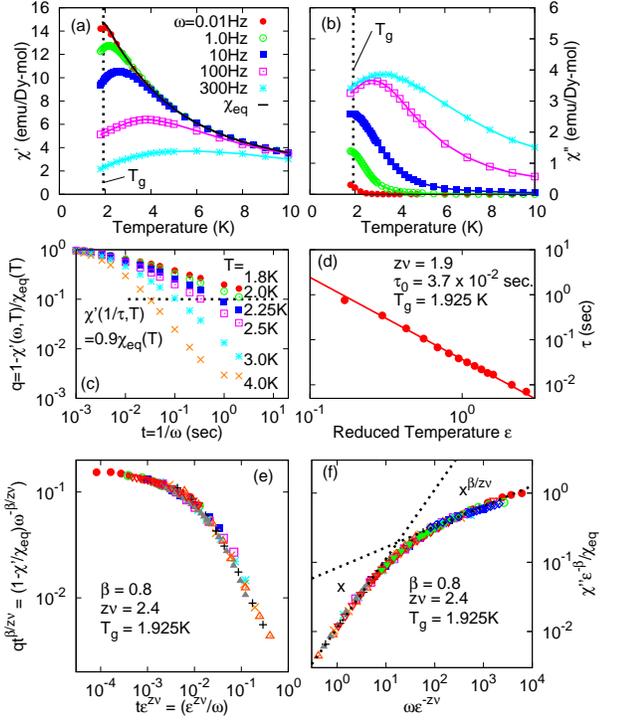}
 \end{center}
 \caption{ \label{fig3}(Color online) (a), (b) $T$ dependences of $\chi'$ (a) and $\chi''$ (b) in zero bias dc field $H$ $=$ 0. The soild line represents the thermal equilibrium susceptibility. (c) Time dependences of $q(t)$ at various $T$ values.  The horizontal dotted line represents the condition for determining $\tau(T)$. (d) Log-log plot of $\tau$ vs $\varepsilon$. (e) and (f) Full scaling plots  of $qt^{\beta/z\nu}$ vs $t\varepsilon^{z\nu}$ and $\chi'' \varepsilon^{-\beta}/\chieq$ vs $\omega\varepsilon^{-z\nu}$, respectively.  Dotted lines in (f) represent the asymptotic behaviors of the scaling function $G(x)$ as $x$ for $x$ $\rightarrow$ 0 and $x^{\beta/z\nu}$ for $x$ $\rightarrow$ $\infty$.
 }
\end{figure}

Next, we present the dynamical features of the SG transition in \dyyrusi\ in zero bias dc field. Figures \ref{fig3}(a) and 3(b) show the $T$ dependences of the real and imaginary parts of the ac susceptibility, respectively. In Fig. \ref{fig3}(a), the thermodynamic equilibrium susceptibility $\chieq$ is also shown, which is determined as $\chieq (T)$ $=$ $\mbox{d}\Meq (T,H)/\mbox{d}H \mid _{H \rightarrow 0}$ from the dc magnetization under the FC condition. The real part of the ac susceptibility $\chi'$ deviates from $\chieq$ at low temperature, and correspondingly, the imaginary part $\chi''$ becomes finite. The observed dynamics is much slower than that of the conventional SGs \cite{SGreview}. Firstly, it originates from the long microscopic relaxation time, being estimated at $\tau_{0} \, = \, 3.7 \times 10^{-2}$ sec by the following dynamics scaling analysis. The strong Ising anisotropy, and the consequent high energy barrier of the spin flipping, is responsible for the long microscopic relaxation time. A similar slow dynamics was observed in several Ising magnets \cite{SpinIceMatsuhira,QuilliamLiHoYF}. On the other hand, we did not find such slow dynamics in a nearly isotropic SG \gdyrusi , which is similar SG material to \dyyrusi\ except for its strength of the magnetic anisotropy. The enormous slowing down observed near $\Tg$ should be the critical slowing down of the SG phase transition, as described later. 

Here, we examine the dynamic scaling analyses of the ac susceptibility.  According to the dynamic scaling hypothesis, the characteristic relaxation time $\tau$ exhibits the critical divergence as $\tau \, = \, \tau_{0} \varepsilon ^{-z\nu}$, where $z$ and $\nu$ are the dynamic and correlation length critical exponents, respectively. In this study, we identify $\tau (T)$ using the criterion of $\chi' (1/\tau;T)$ $=$ $0.9 \chieq (T)$. A similar criterion was also used to determine $\tau$ in Ref. 7. We plot $q(t;T)$ $=$ $1-\chi' (\omega; T)/\chieq(T)$ with $t$ $=$ $1/\omega$, corresponding to the dynamic spin correlation function, against time at several temperatures in Fig. \ref{fig3}(c). The criterion for identifying $\tau$, $q(\tau(T);T)$ $=$ $0.1$, is represented by the horizontal dotted line in the figure. Because the frequency dependence of the ac susceptibility of \dyyrusi\ is very strong, as shown in Fig. \ref{fig3}(a), we can see the decay of the spin correlations over three decades in magnitude at 4 K in Fig. \ref{fig3}(c). As temperature decreases, the dynamics becomes slower, and $q(t)$ does not reach 0.1 below $\Tg$ $=$ 1.925 K. This result is consistent with the general feature of $\tau$, being infinite at $\Tg$, and hence, we concluded that our criterion for identifying $\tau$ is appropriate for the present compound. Figure \ref{fig3}(d) shows a log-log plot of $\tau (T)$ vs $\varepsilon$. In this plot, $\Tg$ was assigned as 1.925 K, being the same value obtained by the static critical scaling analysis, and we obtained the exponent $z\nu$ $=$ 1.9.  Furthermore, we show the full dynamic scaling plots of $q(t;T)$ and $\chi''(\omega;T)$ in Figs. \ref{fig3}(e) and 3(f), respectively.
Both $q(t;T)$ and $\chi''(\omega;T)$ in the ranges of $\varepsilon$ $\leq$ 2.0 and $\omega$ $\leq$ $1000$ Hz are appropriately scaled with $\beta$ $=$ $0.8 \pm 0.1$, $z\nu$ $=$ $2.4 \pm 0.2$, and $\Tg$ $=$ 1.925 K, in the following forms of   
\begin{align}
q(t;T) &= t^{-\beta/z\nu} Q\left( t\varepsilon^{z\nu} \right), \notag \\ 
\chi''(\omega;T)/\chieq (T) &= \varepsilon^{\beta} G\left(\omega\varepsilon^{-z\nu}\right) .
\end{align}
The obtained critical exponents are roughly consistent with those obtained by the static scaling analysis and the analysis of $\tau(T)$. Surprisingly, the value of $z\nu$ $(\approx \, 2)$ is much smaller than that in other SGs, for instance, $z\nu$ $\approx$ 11 in \femntio\ \cite{NordbladFeMnTiO,JonssonFeMnTiO}. On the other hand, this value is consistent with that predicted in the mean-field theory, where $\znuMF$ is 2 with $\zMF$ $=$ 4 and $\nuMF$ $=$ 1/2 \cite{SGreview}. 

\begin{figure}[ht]
 \begin{center}
 \includegraphics[clip,width=8cm]{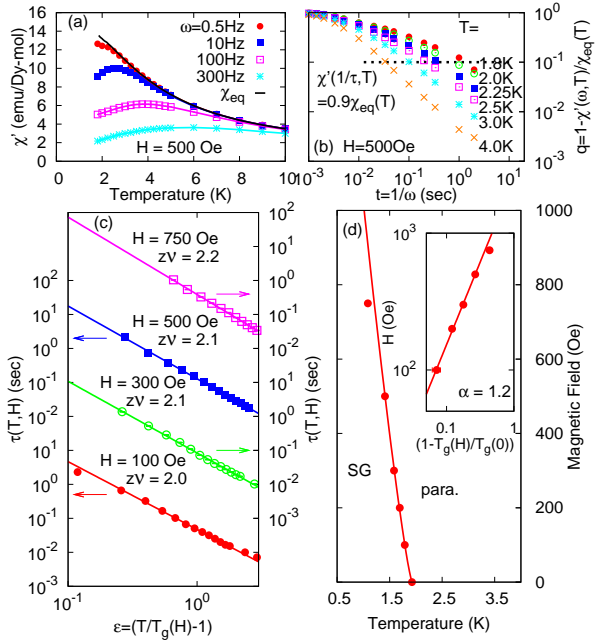}
 \end{center}
 \caption{ \label{fig4}(Color online) (a) $T$ dependences of $\chi'$ in a dc bias field $H$ $=$ 500 Oe. The soild line represents the thermal equilibrium susceptibility. (b) Time dependences of $q(t;T,H)$ at various $T$ values in $H$ $=$ 500 Oe.  The horizontal dotted line represents the condition for determining $\tau(T,H)$. (c) Log-log plots of $\tau(T,H)$ vs $\varepsilon$ at various dc bias fields. (d) $H$ dependence of the SG transition temperature $\Tg(H)$. The inset shows  a log-log plot of $H$ vs $(1-\Tg(H)/\Tg(0))$. 
}
\end{figure}

The static and dynamic critical exponents $\gamma$, $\beta$, $\delta$, and $z\nu$, and the cusp-like anomaly of the specific heat at $\Tg$ indicate that the SG transition in \dyyrusi\ belongs to the mean-field universality class. Hence, the emergence of the RSB is naturally expected. To address this, we examine the stability of the SG phase in finite magnetic field through the ac susceptibility measurements. The SG phase transition in magnetic field was mostly addressed by observing the irreversibility of magnetization; however, it may not be evidence of the thermal equilibrium SG phase transition alone because of  the existence of the nonequilibrium slow dynamics in SG materials \cite{IrreversibilityM}. A better way to determine the SG transition temperature in field $\Tg(H)$ is the analysis of the $T$ dependence of the characteristic relaxation time in field $\tau (T,H)$ to examine whether $\tau(T,H)$ obeys the dynamic critical divergence represented by
\begin{equation}
\tau(T,H) = \tau_{0}(H) |T/\Tg(H) -1|^{-z\nu} 
\label{eq3}
\end{equation}
at $T$ $>$ $\Tg(H)$. If the SG state is stable even in finite field, as well as in zero field, the critical divergence of $\tau(T,H)$ is observed. Otherwise, the experimental data does not exhibit the critical divergent behavior except for the case in zero field. In \femntio , the absence of SG in finite field was addressed by this way \cite{NordbladFeMnTiO,JonssonFeMnTiO}. 

Figure \ref{fig4}(a) shows the $T$ dependence of the real part of the ac susceptibility $\chi'(\omega; T,H)$ at a dc bias field of $H$ $=$ 500 Oe. The thermal equilibrium susceptibility $\chieq(T,H)$ $=$  $\mbox{d}\Meq (T,H)/\mbox{d}H \mid _{H = 500 \mbox{\scriptsize Oe}}$ is also shown. In Fig. \ref{fig4}(b), $q(t=1/\omega;T,H)$ $=$ $1-\chi' (\omega; T, H)/\chieq(T,H)$ at $H$ $=$ 500 Oe is shown. The same criterion for determining $\tau (T,H)$ as in zero field, $q(\tau(T,H);T,H)$ $=$ 0.1, is adopted, which is denoted by the horizontal dotted line in the figure. This criterion is employed and the data set of $\tau(T,H)$ is obtained at each dc bias field of $H$ $\leq$ $1000$ Oe. The log-log plots of $\tau(T,H)$ vs $T/\Tg(H) -1$ at several fields are presented in Fig. \ref{fig4}(c). In any field, $\tau(T,H)$ obeys the critical divergent behavior represented by eq. (\ref{eq3}) with $z\nu$ $\approx$ $2.0$. The value of $z\nu$ is almost $H$-independent, indicating that the SG phase transitions in both zero and finite fields belong to the same universality class, that is, the mean-field one. The $H$ dependence of $\Tg(H)$ shown in Fig. \ref{fig4}(d) exhibits the AT-like behavior, being represented as 
\begin{equation}
H = A \left[ 1 - \Tg(H)/\Tg(0) \right]^{\alpha}
\end{equation}
with $\alpha$ $=$ 1.2 and $A$ $=$ 2500 Oe. The obtained $\alpha$ and $A$ are compatible with the calculated values based on the SK model of $\alpha$ $=$ 1.5 and $A$ $=$ 3300 Oe \cite{AT}.  These results indicate the presence of the SG phase transition in finite field and the RSB SG state in \dyyrusi\ in contrast to the short-range Ising SG \femntio  . 


It is not well understood why the SG transition in \dyyrusi\ belongs to the mean-field universality class. The compound belongs to the class of the 3-dimensional Ising SG  with the long-range RKKY interaction that decays with distance as $\sim$ $r^{-3}$. According to the theoretical arguments on the $d$-dimensional SG with the long-range interaction represented as $r^{-\sigma}$, the $\sigma$ $=$ $d$ systems below the upper critical dimension, corresponding to the 3-dimensional RKKY interaction systems, belong to the non-mean-field universality class \cite{LRSGRG}. In a similar Ising system, such as \lihoyf , where the long-range dipolar interaction decaying as $r^{-3}$ is dominant, the presence of the SG phase is still controversial experimentally \cite{JonssonLiHoYF,QuilliamLiHoYF} and a non-mean-field-type SG transition was predicted numerically \cite{TamDipolar}. Further experimental, numerical and theoretical investigations for the long-range SG are required. 

In conclusion, we have investigated the SG phase transition in the long-range RKKY Ising SG \dyyrusi\   and observed the mean-field critical phenomena. The experimental results show the presence of the SG phase under finite field and the validity of the SK model in \dyyrusi , strongly suggesting that the RSB SG state emerges in real materials with the long-range interaction as well as in the artificial infinite interaction system. 


\end{document}